\documentclass[12pt]{article}
\usepackage{amsmath, amsthm, amscd, amsfonts, amssymb, graphicx, color}
\usepackage[bookmarksnumbered, colorlinks, plainpages]{hyperref}

\theoremstyle{definition}

\theoremstyle{remark}

\numberwithin{equation}{section}

\usepackage{cite}
\usepackage{geometry}
\usepackage{caption}
\usepackage{subcaption}
\usepackage{hyperref}
\hypersetup{
        colorlinks=true,
        citecolor=black,
        linkcolor=black,
        urlcolor=black
    }

\begin{document}

\title{Interpret the estimand framework from a causal inference perspective}
\author{Jinghong Zeng$^{1, 2}$\\
\\
\small $^1$Department of Statistics, University of Auckland, New Zealand\\
\small Email: jzen696@aucklanduni.ac.nz\\
\small $^2$Department of Statistics and Programming, Jiangsu Hengrui Pharmaceuticals Co. Ltd., China}
\date{}
 \maketitle

\begin{abstract}

The estimand framework proposed by ICH in 2017 has brought fundamental changes in the pharmaceutical industry. It clearly describes how a treatment effect in a clinical question should be precisely defined and estimated, through attributes including treatments, endpoints and intercurrent events. However, ideas around the estimand framework are commonly in text, and different interpretations on this framework may exist. This article aims to interpret the estimand framework through its underlying theories, the causal inference framework based on potential outcomes. The statistical origin and formula of an estimand is given through the causal inference framework, with all attributes translated into statistical terms. We describe how five strategies proposed by ICH to analyze intercurrent events are incorporated in the statistical formula of an estimand, and we also suggest a new strategy to analyze intercurrent events. The roles of target populations and analysis sets in the estimand framework are compared and discussed based on the statistical formula of an estimand. This article recommends continuing studying causal inference theories behind the estimand framework and improving the estimand framework with greater methodological comprehensibility and availability.

\end{abstract}

\section{Introduction}

The estimand framework was drafted in 2017 by ICH in Efficacy Guideline E9(R1) as addendum to Efficacy Guideline E9 and later published in 2019 \cite{ich_expert_working_group_statistical_1998, ich_expert_working_group_addendum_2019}. It has gained increasing attention in the pharmaceutical industry and the academia \cite{pohl_estimand_2021, kahan_estimand_2024, heinrich_estimand_2025, drury_estimand_2025, lanius_estimand_2025}. Many clinical trials have used the estimand framework to develop new drugs for both oncology and non-oncology diseases, and professional working groups such as the Oncology Estimand Working Group have been initiated to study how the estimand framework should be incorporated in pharmaceutical practices \cite{onc_estimand_2018}. 

The estimand framework aims to clearly define a clinical question. This means to clearly define a treatment effect in this clinical question. The estimand framework introduces and compares estimands, estimators and estimates with regard to statistical roles in estimation of treatment effects. An estimand is a precise definition of a treatment effect in a clinical question, an estimator is a statistical method that estimates this estimand, and an estimate is a result from this estimator \cite{ich_expert_working_group_addendum_2019}. Five attributes are proposed for an estimand. They are treatments, endpoints, a target population, intercurrent events, and a population-level summary \cite{ich_expert_working_group_addendum_2019}. Generally, Treatments are medical interventions that patients would take in clinical trials. Endpoints are outcomes used to assess efficacy and safety of treatments. A target population is a group of patients with medical conditions of clinical interest. Intercurrent events are events that happen after treatment initiation and affect the definition of a treatment effect. A population-level summary is an approach to estimate the treatment effect in a target population.  

Intercurrent events are frequent in practice but conceptually novel. E9(R1) listed many examples for intercurrent events, such as use of concomitant therapies, treatment switching and death before endpoint measurement. To suit different study objectives, five strategies have been proposed to reduce bias from intercurrent events. They are the treatment policy strategy, the hypothetical strategy, the composite variable strategy, the while on treatment strategy, and the principal stratum strategy \cite{ich_expert_working_group_addendum_2019}.

The estimand framework is still a relatively new concept, where different interpretations may exist. Any contribution to further clarification of the estimand framework would be beneficial for clinical development. The original ideas in the estimand framework are mainly conveyed through text, but a statistical description of the estimand framework may be more intuitive, precise and holistic. The core of the estimand framework is treatment effects, and attributes of an estimand are developed to define a treatment effect. It would be helpful to understand a treatment effect through a statistical causal inference framework and see how these attributes and strategies are related to underlying theories.

\section{The causal inference framework for estimands}

\begin{figure}[!htbp]
     \centering
\includegraphics[width=0.7\textwidth]{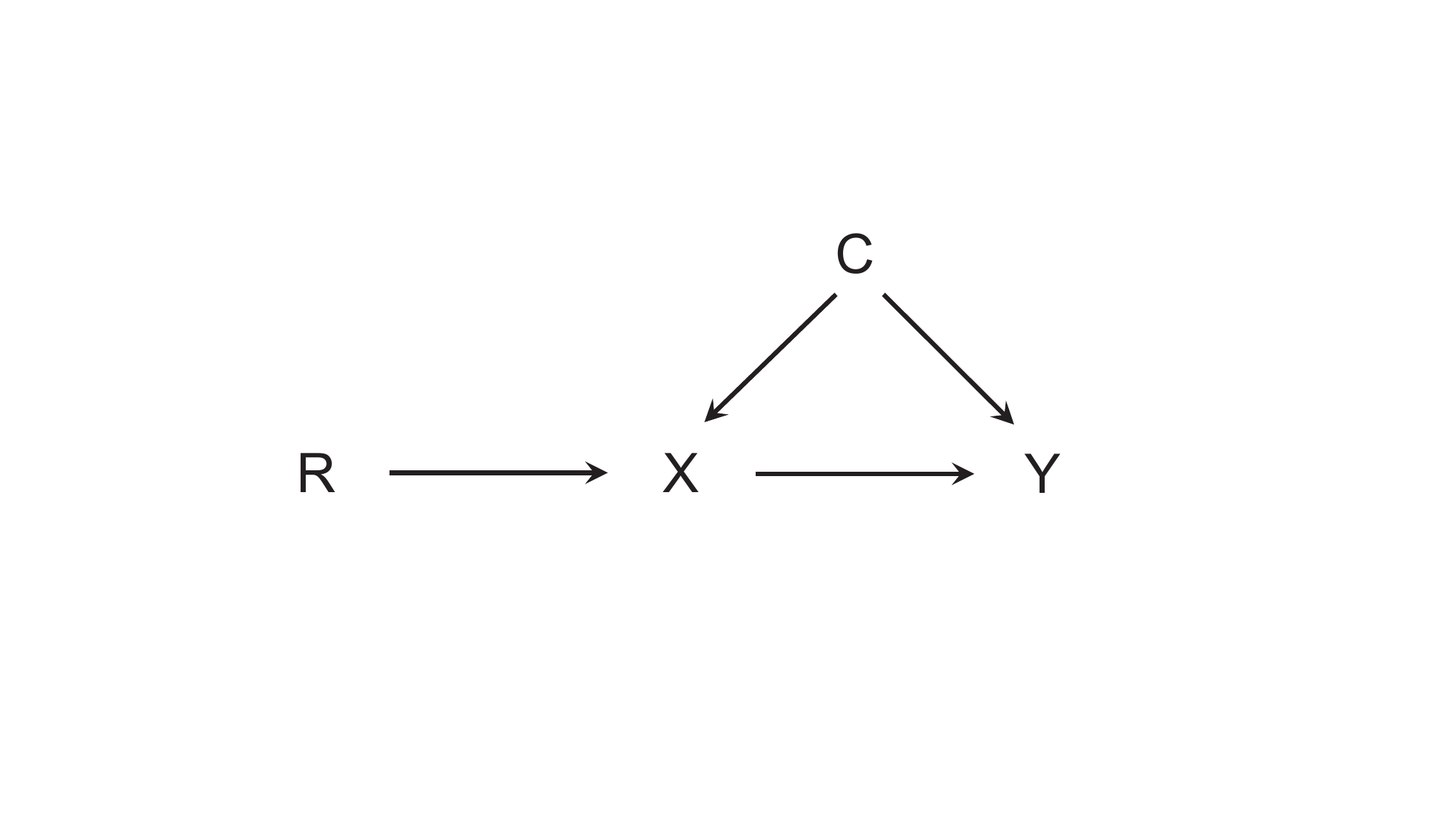}
        \caption{Causal directed acyclic graph for $X$, $Y$, $R$, $C$.}
        \label{fig:cdag1}
\end{figure}

A causal inference framework is based on the potential outcome framework \cite{donald_b_rubin_estimating_1974, paul_w_holland_statistics_1986, imbens_local_1994}. Suppose there is an ideal two-arm randomized controlled clinical trial, with full compliance to treatment and no intercurrent events. This clinical trial has a sample size of $N$, a binary treatment $X$, a continuous endpoint $Y$, a randomization scheme $R$ and some confounders $C$ that affect both $X$ and $Y$. $X$, $Y$, $R$ are random vectors of length $N$, and $C$ is a random matrix of row dimension $N$. $X_i$, $Y_i$, $R_i$, $C_i$ represent random variables or vectors for the participant $i$, where $i \in \{1, 2, 3, \ldots, N\}$. $R_i=0$ means that the participant is assigned to the control arm, and $R_i=1$ means that the participant is assigned to the treatment arm. $X_i(R_i=0)=0$ means that the participant would take the control treatment if assigned to the control arm, and $X_i(R_i=1)=1$ means that the participant would take the experimental treatment that is of primary clinical interest if assigned to the treatment arm. $Y_i(X_i(R_i=0)=0)$ would be the endpoint if the participant took the control treatment as assigned, and $Y_i(X_i(R_i=1)=1)$ would be the endpoint if the participant took the experimental treatment as assigned. $R_i$, $X_i$ and $Y_i$ are potential outcomes. They are not real because we hypothesize what would happen if the participant took the control treatment or instead took the experimental treatment.  Figure \ref{fig:cdag1} is a causal directed acyclic graph that shows the causal relationships between $X$, $Y$, $R$, $C$. $R$ affects $Y$ only through $X$.

A clinical question is to estimate the overall treatment effect of taking the experimental treatment versus taking the control treatment on the endpoint over all participants. For the participant $i$, if he had $X_i(R_i=0)=0$ then he would have $Y_i(X_i(R_i=0)=0)$, and if he had $X_i(R_i=1)=1$ then he would have $Y_i(X_i(R_i=1)=1)$. For this participant, the individual treatment effect (ITE) is the difference between two potential outcomes of $Y_i$. That is,
 \begin{eqnarray}
 ITE &=& Y_i(X_i(R_i=1)=1) - Y_i(X_i(R_i=0)=0). \nonumber
 \end{eqnarray}

This individual treatment effect controls confounders on the endpoint within the same participant and means how the endpoint would change when only the treatment condition changes. There are $N$ individual treatment effects. They can be assumed to be equal or unequal. Suppose we are interested in an average treatment effect (ATE) among all participants. The ATE is an average of all individual treatment effects and can be described as 
 \begin{eqnarray}
ATE &=& E(Y(X(R=1)=1) - Y(X(R=0)=0)~|~C). \nonumber
 \end{eqnarray}

In the clinical question, the estimand is the ATE. If there is no confounder, the ATE would become $E(Y(X(R=1)=1) - Y(X(R=0)=0))$. Without adjusting for existing confounding effects, the ATE estimation may be biased.

Further, based on expectation properties, we have 
 \begin{eqnarray}
ATE &=& E(Y(X(R=1)=1)~|~C)-E(Y(X(R=0)=0)~|~C). \nonumber
 \end{eqnarray}

This also means that the estimand is the difference between the average treatment effect of taking the experimental treatment among all participants and the average treatment effect of taking the control treatment among all participants. The problem is that, in the real world, each participant only takes one kind of treatment, and only one of the two hypothetical situations with regard to two arms can happen. In this clinical trial setting, for each participant, the observed randomized assignment $R_i^o$ is either 0 and 1, the observed treatment $X_i^o$ is either $X_i(R_i=0)$ or $X_i(R_i=1)$, and the observed outcome $Y_i^o(X_i^o)$ is either $Y_i(X_i(R_i=0))$ or $Y_i(X_i(R_i=1))$.  The relationship between potential outcomes and observed variables can be described as
 \begin{eqnarray}
X_i^o &=& X_i(R_i=R_i^o), \nonumber \\
Y_i^o &=& Y_i(X_i(R_i=R_i^o)). \nonumber
 \end{eqnarray}
 
What we can estimate from the actual clinical trial data is the difference (D) between the average treatment effect from participants who take the experimental treatment in the treatment arm and the average treatment effect from participants who take the control treatment in the control arm. That is,
 \begin{eqnarray}
D &=& E(Y^o~|~X^o=1, R^o=1, C)-E(Y^o~|~X^o=0, R^o=0, C). \nonumber
 \end{eqnarray}

With certain assumptions, including the stable unit treatment value assumption, the randomization assumption and the exclusion restriction assumption, it can be proved that D equals to the ATE, through
 \begin{eqnarray}
D &=& E(Y^o~|~X^o=1, C)-E(Y^o~|~X^o=0, C), \nonumber \\
 &=& E(Y(X(R=1)=1)~|~C)-E(Y(X(R=0)=0)~|~C). 
  \label{eqn:eqn-d}
 \end{eqnarray}

Suppose we choose a causal linear model to estimate the estimand. The linear model is
 \begin{eqnarray}
Y^o = \beta_0 + \beta_1 X^o + \beta_2 C + \varepsilon,  \nonumber \\
E(\varepsilon)=0, Var(\varepsilon) = \sigma^2,  \nonumber
 \end{eqnarray}
 
where $E(Y^o ~|~X^o, C)=\beta_0 + \beta_1 X^o + \beta_2 C$. Then, we have two equations with different observed treatment levels as
 \begin{eqnarray}
E(Y^o ~|~X^o=0, C) &=& \beta_0 + \beta_2 C, \nonumber \\
E(Y^o ~|~X^o=1, C) &=& \beta_0 + \beta_1 + \beta_2 C, \nonumber
 \end{eqnarray}

which implies that 
 \begin{eqnarray}
\beta_1 = E(Y^o ~|~X^o=1, C)-E(Y^o ~|~X^o=0, C),  
  \label{eqn:eqn-beta1}
 \end{eqnarray}

Hence, $\beta_1$ equals to the ATE through equations \ref{eqn:eqn-d} and \ref{eqn:eqn-beta1}. It is an estimator to the estimand from this linear model. After the linear model is built with the actual clinical trial data, an estimate $\hat \beta_1$ on $\beta_1$  would be obtained. $\hat \beta_1$ is an estimate to the estimand.

The statistical formula of the estimand as $E(Y(X(R=1)=1) - Y(X(R=0)=0)~|~C)$ has already contained five attributes proposed in the estimand framework. The attribute of treatments is represented by $X$. The attribute of endpoints is represented by $Y$. The attribute of a population-level summary is represented by the difference in the average treatment effect between taking the experimental treatment and taking the control treatment among all participants. The attribute of intercurrent events is not necessary in this clinical trial setting because no intercurrent events are assumed. The attribute of a target population is implicitly stated as the clinical trial population and can be made explicit in the definition of the estimand. We can introduce a selection variable $S$ that indicates how the target population is selected from the general public and update the estimand to $E(Y(X(R=1)=1) - Y(X(R=0)=0)~|~C, S)$. 

The causal inference framework is not restricted to the above situation. It is applicable to multivalued treatments, discrete outcomes, nonlinear risk comparisons and Bayesian methods. It is also the underlying theory that applies to any effort of estimating treatment effects. We may have used it, even if the causal inference framework does not appear explicitly in inference, because potential outcomes have been transformed to observed variables in statistical models and we only see observed variables. Causal inference is a mature, continuing research area in Statistics. Methods may be available for difficult practical problems. Hence, it would be recommended that we try to interpret the estimand from a causal inference perspective based on our study objectives, and understand or choose relevant methods to estimate estimands.

\section{Causal inference with intercurrent events}
\begin{figure}[!htbp]
     \centering
\includegraphics[width=0.9\textwidth]{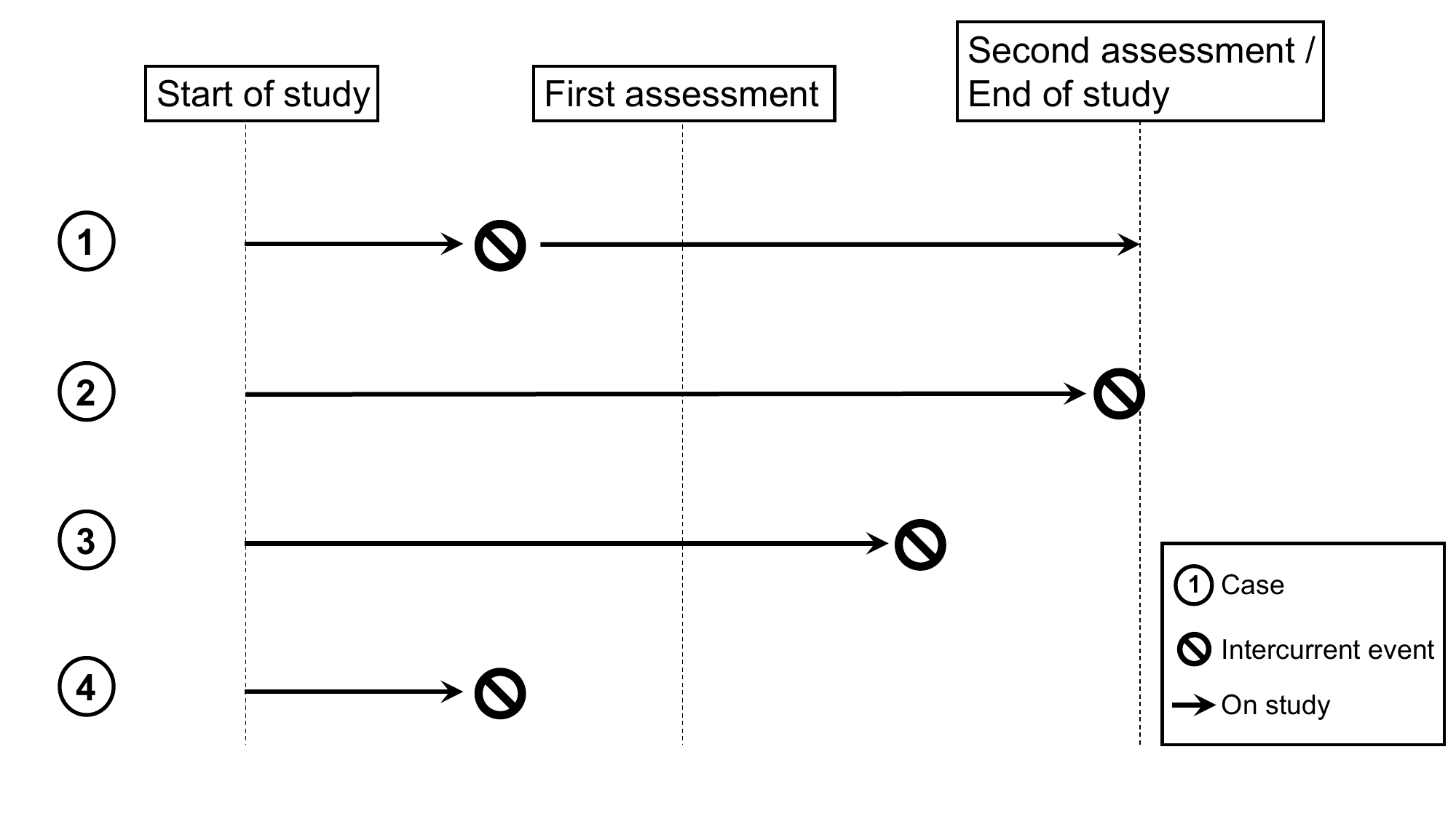}
        \caption{Different intercurrent events. Case (1): Participants use concomitant therapies after treatment initiation and continue study till the end. The second assessment on the endpoint is available. Case (2): Participants die by the second assessment. The second assessment on the endpoint is not available. Case (3): Participants die between two assessments. The second assessment on the endpoint is not available but the first assessment on the endpoint is available. Case (4): Participants discontinue treatment after treatment initiation due to treatment toxicity and withdraw from study. No endpoint assessment is available.}
        \label{fig:inter2}
\end{figure}

The attribute of intercurrent events can also be incorporated in the statistical formula of an estimand. Following the clinical trial example in the last section, suppose now the clinical trial has intercurrent events as shown in figure \ref{fig:inter2}. Also suppose the endpoint is not related to death and the second assessment on the endpoint is used for primary analysis. We use these intercurrent events to discuss the statistical formula of an estimand in different strategies.

\subsection{The treatment policy strategy}

The treatment policy strategy is to include intercurrent events in the treatment definition. In case (1), when concomitant therapies are used, the endpoint may also be affected by concomitant therapies in addition to the control and experimental treatments. Hence, we cannot estimate the ATE of the experimental treatment versus the control treatment on the endpoint without adjusting for concomitant therapies. Through the treatment policy strategy, we re-define $X$ as the control or experimental treatment plus any concomitant therapy. Then we can use all participants from case (1) in data analysis. The estimand formula is still $E(Y(X(R=1)=1) - Y(X(R=0)=0)~|~C)$, but now the estimand has been changed into the ATE of the experimental treatment plus any concomitant therapy versus the control treatment plus any concomitant therapy.

This strategy does not estimate the pure effect of the experimental treatment. When concomitant therapies are common for use with the experimental treatment in the real world, this strategy would be appropriate.

\subsection{The hypothetical strategy}

The hypothetical strategy is to hypothesize non-existence of intercurrent events and make relevant endpoints missing. Suitable statistical methods should be used to impute missing endpoints.

In case (1), if concomitant therapies are prohibited by the clinical trial protocol, we cannot include concomitant therapies in the treatment definition and the endpoint assessments after prohibited therapies are used are not appropriate for analysis. Through the hypothetical strategy, we make the second assessment on the endpoint after prohibited therapies are used missing and impute missing endpoint assessments as if there were only the control and experimental treatments.  In case (2), the second assessment on the endpoint is naturally missing due to death. Through the hypothetical strategy, we impute missing endpoint assessments with the assumption that participants are alive at this time.  For both cases, the estimand formula is still $E(Y(X(R=1)=1) - Y(X(R=0)=0)~|~C)$, and the meaning of the estimand is not changed.

This strategy estimates the pure effect of the experimental treatment. When the pure treatment effect is of major clinical importance in the real world, this strategy would be appropriate.

\subsection{The composite variable strategy}

The composite variable strategy is to include intercurrent events in the endpoint definition.  In case (2), suppose the endpoint is a disability index measured by questionnaires and it does not consider death. Since participants die by the second assessment, the questionnaire measurement is not done at the second assessment. However, death already indicates no physical function. Through the composite variable strategy, death is included in the worst disability category and thus can be used in data analysis. The estimand formula is still $E(Y(X(R=1)=1) - Y(X(R=0)=0)~|~C)$, but now the definition of $Y$ has been changed from disability measurement with no death included to disability measurement with death included in the worst disability category.

This strategy estimates the pure effect of the experimental treatment on a different endpoint. When composite endpoints are realistic in the real world, this strategy would be appropriate.

\subsection{The while on treatment strategy}

The while on treatment strategy is to use available endpoints measured before intercurrent events and discard any clinical trial data afterwards. In case (3), participants survive the first assessment, but the endpoint is not measured at the second assessment due to death. Through the while on treatment strategy, we use the first assessment on the endpoint and all data up to the first assessment when participants are included in primary analysis. This means that, for participants with intercurrent events, we only use their data that are recorded during treatment before intercurrent events. The estimand formula is still $E(Y(X(R=1)=1) - Y(X(R=0)=0)~|~C)$, but now the definition of $Y$ has been changed from measurement at the second assessment to measurement at the first or second assessment.

This strategy also estimates the pure effect of the experimental treatment on a different endpoint. When time is well adjusted in the endpoint definition, such as multiple-fixed-time endpoints as in this example and time-to-event endpoints, this strategy would be appropriate, otherwise we may have to explicitly adjust for time in statistical models.

\subsection{The principal stratum strategy}

The principal stratum strategy is to identify a subpopulation of participants who would or would not experience intercurrent events and estimate the ATE in this subpopulation instead of the entire target population.

In case (4), suppose we only want to know the treatment effect in the participants who would tolerate treatment toxicity. Tolerability of treatment toxicity is an intrinsic characteristic and should be interpreted from a hypothetical perspective. Suppose tolerability is represented by $T$. $T=0$ represents not tolerating treatment toxicity, and $T=1$ represents tolerating treatment toxicity. Now, there are four subpopulations represented by $P$. $P=1$ are participants who would tolerate both the control and experimental treatments if they took both treatments. $P=2$ are participants who would tolerate neither the control treatment nor the experimental treatment if they took both treatments. $P=3$ are participants who would tolerate the control treatment and would not tolerate the experimental treatment if they took both treatments. $P=4$ are participants who would not tolerate the control treatment and would tolerate the experimental treatment if they took both treatments. And the subpopulation of participants who would tolerate both treatments, that is $P=1$, becomes our new target population. These four subpopulations may not be directly observed in the real world and may have to be estimated from data. They are different from subgroups of participants who tolerate or do not tolerate treatment toxicity in the real world. For example, there is a subgroup of participants in the control arm who tolerate the control treatment. From this subgroup, some participants might tolerate the experimental treatment if assigned to the treatment arm instead, while the others might not tolerate the experimental treatment if assigned to the treatment arm instead. Hence, this subgroup may be a mix of $P=1$ and $P=3$.

Potential outcomes should be modified to take $T$ into account. For example, $Y_i(X_i(R_i=0)=0, T_i=0)$ would be the outcome if the participant took but did not tolerate the control treatment.
$Y_i(X_i(R_i=1)=1, T_i=0)$ would be the outcome if the participant took but did not tolerate the experimental treatment.
$Y_i(X_i(R_i=0)=0, T_i=1)$ would be the outcome if the participant took and tolerated the control treatment.
$Y_i(X_i(R_i=1)=1, T_i=1)$ would be the outcome if the participant took and tolerated the experimental treatment. The estimand formula now becomes $E(Y(X(R=1)=1) - Y(X(R=0)=0)~|~C, P=1)$, with no changes in the definitions of $X$ and $Y$.  

This strategy estimates the pure treatment effect in a different target population. When specific subpopulations are of clinical interest, this strategy would be appropriate. Statistical methods with regard to this strategy may be complex. Bayesian inference would be a good choice for this strategy \cite{rubin_bayes_1978, imbens_bayes_1997, jinghong_bayes_2023}.

\subsection{Further possibilities}

The above five strategies proposed in the estimand framework can be used in a flexible way. Several strategies can be used together. For example, the treatment policy strategy to analyze concomitant therapies can be used with the principal stratum strategy to identify a subpopulation. A composite strategy has been proposed by industrial professionals that the treatment policy strategy to analyze allowed concomitant therapies is combined with the hypothetical strategy to analyze prohibited concomitant therapies. 

\begin{figure}[!htbp]
     \centering
\includegraphics[width=0.7\textwidth]{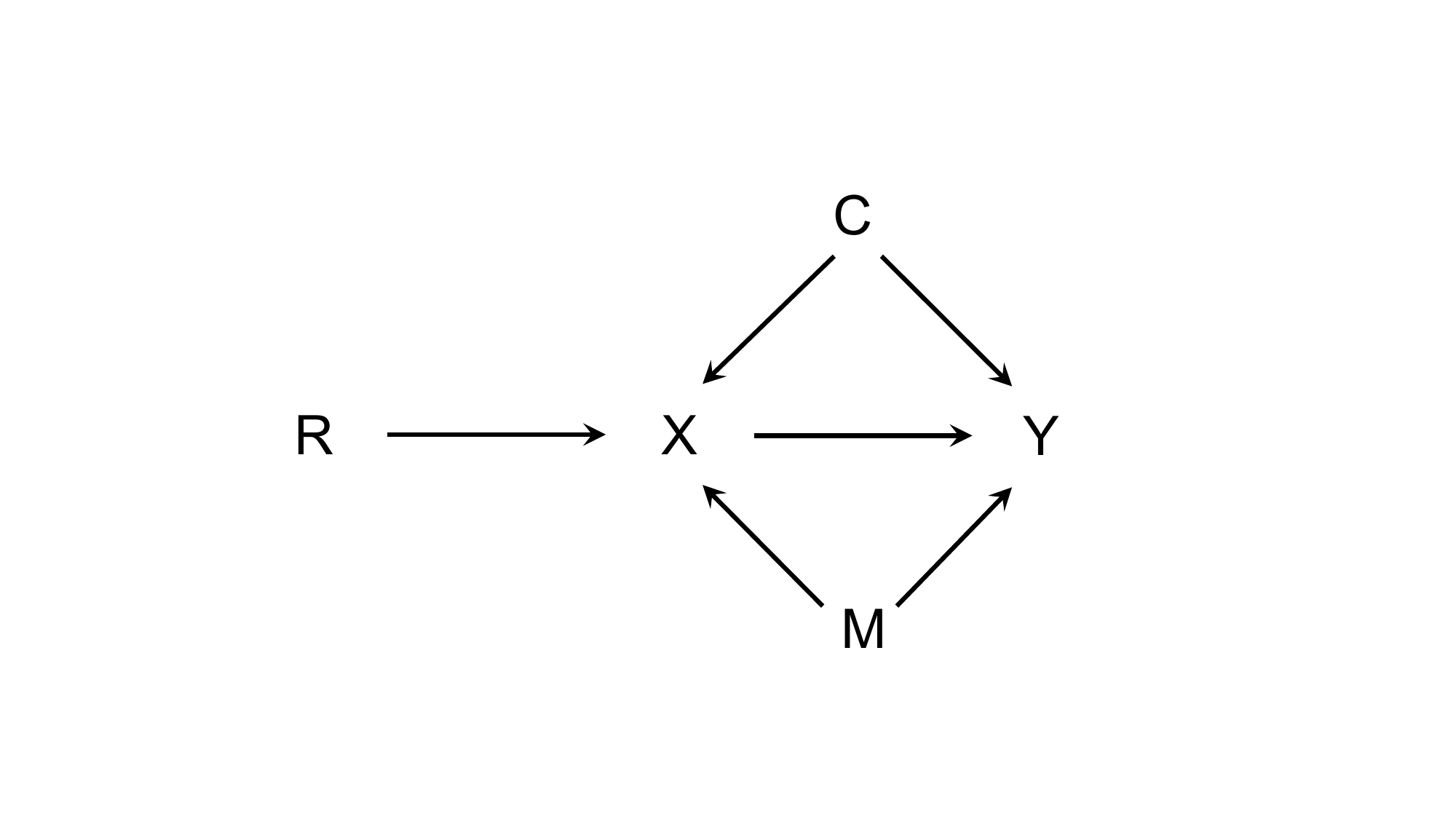}
        \caption{Causal directed acyclic graph for $X$, $Y$, $R$, $C$, $M$.}
        \label{fig:cdag2}
\end{figure}

In addition to the five strategies, there are also possible other strategies to analyze intercurrent events. For example, in case (1), suppose the dose of the control or experimental treatment has to be reduced after concomitant therapies are used, and concomitant therapies affect the endpoint. We define a new variable $M$ to represent concomitant therapies. $M$ may be a binary variable that indicates whether concomitant therapies are used or not. It can also be the total dose or the dose frequency of concomitant therapies. Now, $M$ becomes a new confounder that affects both $X$ and $Y$. Figure \ref{fig:cdag2} is a causal directed acyclic graph that shows the causal relationships between $X$, $Y$, $R$, $C$, $M$.

The estimand formula now should be conditional on $M$ as $E(Y(X(R=1)=1) - Y(X(R=0)=0)~|~C, M)$. Hence, the causal linear model would become $Y^o = \beta_0 + \beta_1 X^o + \beta_2 C + \beta_3 M+ \varepsilon$, where $\beta_1$ equals to the ATE after adjusting for new confounding effects. 

This new strategy defines intercurrent events as confounders and uses model adjustment. Compared to the treatment policy strategy and the hypothetical strategy, the model adjustment strategy does not modify the definitions of $X$, $Y$ and the target population, and it does not modify any clinical trial data. If the confounding effects from intercurrent events can be well adjusted for, the model adjustment strategy would be appropriate.

In practice, to create new strategies, more aspects should be considered, including efficiency, validity in various applications and suitability for regulatory activities. It would be recommended continuing developing new strategies for intercurrent events beyond the proposal of ICH E9(R1).

\section{Comparison between target populations and analysis sets}

An analysis set is a group of participants that meet specific criteria and is used for specific analytical objectives. ICH E9 proposed different analysis sets, including full analysis set (FAS) and per protocol set (PPS) \cite{ich_expert_working_group_statistical_1998}. There are also other analysis sets, including intention to treat set (ITTS) and safety set (SS).  ITTS is usually all participants enrolled in study. SS comes from ITTS and is usually a group of participants who take treatments at least once. FAS comes from SS and may exclude some participants whose data are not suitable for analysis. PPS comes from FAS and may only include participants whose data have no or little protocol deviation. In practice, definitions of different analysis sets may vary by clinical trial. Some statistical analysis plans may give special considerations to analysis sets based on study design. We may have to interpret analysis sets in the clinical trial context.

Analysis sets are usually subsets of all trial data. Strictly speaking, they are not necessarily the target population of interest. If we build statistical models on data subsets through the estimand framework, we may introduce selection bias and break randomization of the assignment scheme, which means that the participants in an analysis set may not be considered randomized any more. Without the randomization assumption, statistical inference on estimands may be unviable or biased. Hence, we should carefully examine randomization in analysis sets if we want to use them as target populations. On the other hand, the principal stratum strategy provides an approach to estimate treatment effects in various analysis sets, if it is feasible to define analysis sets into subpopulations.

\section{Conclusion}

The estimand framework is advantageously useful for estimation of causal treatment effects. Learning and strengthening statistical causal inference theories behind the estimand framework would facilitate the implementation and advancement of the estimand framework in the pharmaceutical industry.

\section*{Funding statement}

There is no funding for this work.


\begin{thebibliography}{100} 

\bibitem{ich_expert_working_group_statistical_1998} ICH Expert Working Group. Statistical Principles for Clinical Trials E9. 1998. \href{https://www.ich.org/page/efficacy-guidelines#9-1}{www.ich.org/page/efficacy-guidelines\#9-1}

\bibitem{ich_expert_working_group_addendum_2019} ICH Expert Working Group. Addendum on Estimands and Sensitivity Analysis in Clinical Trials to the Guideline of Statistical Principles for Clinical Trials E9(R1). 2019. \href{https://www.ich.org/page/efficacy-guidelines#9-2}{www.ich.org/page/efficacy-guidelines\#9-2}

\bibitem{pohl_estimand_2021} Pohl M, Baumann L, Behnisch R, Kirchner M, Krisam J, and Sander A. Estimands—a basic element for clinical trials. Part 29 of a series on evaluation of scientific publications. Dtsch Arztebl Int 2021; 118:883-8. doi: \href{https://doi.org/10.3238/arztebl.m2021.0373}{10.3238/arztebl.m2021.0373}

\bibitem{kahan_estimand_2024} Kahan BC, Hindley J, Edwards M, Cro S, and Morris TP. The estimands framework: a primer on the ICH E9(R1) addendum. BMJ 2024:e076316. doi: \href{https://doi.org/10.1136/bmj-2023-076316}{10.1136/bmj-2023-076316}

\bibitem{heinrich_estimand_2025} Heinrich M, Zagorscak P, Bohn J, Knaevelsrud C, and Schulze L. Using the ICH estimand framework to improve the interpretation of treatment effects in internet interventions. npj Digit Med 2025; 8:535. doi: \href{https://doi.org/10.1038/s41746-025-01936-0}{10.1038/s41746-025-01936-0}

\bibitem{drury_estimand_2025} Drury T, Bartlett JW, Wright D, and Keene ON. The estimand framework and causal inference: complementary not competing paradigms. Pharm Stat 2025; 24:e70035. doi: \href{https://doi.org/10.1002/pst.70035}{10.1002/pst.70035}

\bibitem{lanius_estimand_2025} Lanius V, Glocker B, L{\"o}sch C, Bratton DJ, Callegari F, Wright M, and Rajam{\"a}ki S. Realizing the benefits of the estimand framework when reporting and communicating clinical trial results—some recommendations. Trials 2025; 26:241. doi: \href{https://doi.org/10.1186/s13063-025-08915-6}{10.1186/s13063-025-08915-6}

\bibitem{onc_estimand_2018} Oncology Estimand Working Group. 2018. \href{https://oncoestimand.github.io/oncowg_webpage/docs/}{oncoestimand.github.io/oncowg\_webpage/docs/}

\bibitem{donald_b_rubin_estimating_1974} Rubin DB. Estimating causal effects of treatments in randomized and nonrandomized studies. J Educ Psychol 1974; 66:688-701. doi: \href{https://doi.org/10.1037/h0037350}{10.1037/h0037350}

\bibitem{paul_w_holland_statistics_1986} Holland PW. Statistics and causal inference. J Am Stat Assoc 1986; 81:945-60. doi: \href{https://doi.org/10.1080/01621459.1986.10478354}{10.1080/01621459.1986.10478354}

\bibitem{imbens_local_1994} Imbens GW and Angrist JD. Identification and estimation of local average treatment effects. Econometrica 1994; 62:467. doi: \href{https://doi.org/10.2307/2951620}{10.2307/2951620}

\bibitem{rubin_bayes_1978} Rubin DB. Bayesian inference for causal effects: the role of randomization. Ann Stat 1978; 6:34-58. \href{http://www.jstor.org/stable/2958688}{www.jstor.org/stable/2958688}

\bibitem{imbens_bayes_1997} Imbens GW and Rubin DB. Bayesian inference for causal effects in randomized experiments with noncompliance. Ann Stat 1997; 25. Publisher: Institute of Mathematical Statistics:305-27. \href{http://www.jstor.org/stable/2242722}{www.jstor.org/stable/2242722}

\bibitem{jinghong_bayes_2023} Zeng J. Bayesian inference on average treatment effects in the PreventS trial data in the presence of unmeasured confounding. Master of Science thesis. The University of Auckland 2023. \href{https://hdl.handle.net/2292/63865}{hdl.handle.net/2292/63865}

\end{thebibliography}
\end{document}